\documentclass[aps,prl,twocolumn,groupedaddress,hidelinks]{revtex4-1}

\usepackage{glossaries} 
\newacronym{rb}{RB}{randomized benchmarking}
\newacronym{gst}{GST}{gate set tomography}
\newacronym{qec}{QEC}{quantum error correction}
\newacronym{se}{SE}{spin echo}
\newacronym{ptm}{PTM}{Pauli transfer matrix}
\newacronym{me}{ME}{Magnus expansion}
\newacronym{dd}{DD}{dynamical decoupling}
\newacronym{dcg}{DCG}{dynamically corrected gate}
\newacronym{ff}{FF}{filter function}
\newacronym{mc}{MC}{Monte Carlo}
\newacronym{cff}{CFF}{correlation filter function}
\newacronym{qft}{QFT}{quantum Fourier transform}
\newacronym{cp}{CP}{completely positive}
\newacronym{povm}{POVM}{positive operator-valued measure}

\usepackage{graphicx,epstopdf,xspace,amssymb,amsmath,float}
\usepackage{qcircuit}
\usepackage{dsfont}
\usepackage[utf8]{inputenc}
\usepackage[english]{babel}
\usepackage{mathtools}
\usepackage{physics}
\usepackage[T1]{fontenc}
\usepackage{csquotes}
\usepackage{xr-hyper}
\usepackage[breaklinks]{hyperref}
\usepackage[nameinlink,capitalise]{cleveref}

\newcommand{\eg}[0]{e.g.\ }

\newcommand{\cf}[0]{c.f.\ }
\newcommand{\mc}[1]{\ensuremath{\mathcal{#1}}}
\newcommand{\mr}[1]{\ensuremath{\mathrm{#1}}}
\newcommand{\e}[0]{\ensuremath{\mr{e}\xspace}}
\renewcommand{\i}[0]{\ensuremath{\mr{i}\xspace}}

\newcommand{\citer}[1]{Ref.~\citenum{#1}}
\newcommand{\citerr}[2]{Refs.~\citenum{#1} and \citenum{#2}}

\newcommand{\fid}[0]{\ensuremath{\mathcal{F}}\xspace}

\newcommand{\infid}[0]{\ensuremath{\mathcal{I}}\xspace}

\newcommand{\oneoverf}{\ensuremath{\flatfrac{1}{f}}\xspace}
\newcommand{\px}[0]{\ensuremath{\sigma_x}\xspace}
\newcommand{\py}[0]{\ensuremath{\sigma_y}\xspace}
\newcommand{\pz}[0]{\ensuremath{\sigma_z}\xspace}
\newcommand{\eye}[0]{\ensuremath{\mathds{1}}\xspace}

\newcommand{\ad}{\ensuremath{^\dagger}\xspace}

\newcommand{\gth}[1]{\ensuremath{^{(#1)}}\xspace}

\newcommand{\Hc}{\ensuremath{H_\mr{c}}\xspace}
\newcommand{\Hn}{\ensuremath{H_\mr{n}}\xspace}
\newcommand{\Hnt}{\ensuremath{\tilde{H}_\mr{n}}\xspace}

\newcommand{\Ue}{\ensuremath{\tilde{U}}\xspace}
\newcommand{\Uc}{\ensuremath{U_\mr{c}}\xspace}
\newcommand{\Pe}{\ensuremath{\tilde{\mathcal{U}}}\xspace}
\newcommand{\Rc}{\ensuremath{\tilde{\mathcal{B}}}\xspace}
\newcommand{\Ba}{\ensuremath{B_\alpha}\xspace}  % noise operators
\newcommand{\Bat}{\ensuremath{\tilde{B}_\alpha}\xspace}  % interaction picture noise operators
\newcommand{\Qc}{\ensuremath{\mathcal{Q}}\xspace}

\newcommand{\decayamps}[0]{\ensuremath{\Gamma}\xspace}

\begin{document}

\title{Filter Functions for Quantum Processes under Correlated Noise}

\author{Pascal Cerfontaine}
\email[]{pascal.cerfontaine@rwth-aachen.de}
\author{Tobias Hangleiter}
\author{Hendrik Bluhm}
\affiliation{JARA-FIT Institute for Quantum Information, Forschungszentrum J\"ulich GmbH and RWTH Aachen University, 52074 Aachen, Germany}


\begin{abstract}
Many qubit implementations are afflicted by correlated noise not captured by standard theoretical tools that are based on Markov approximations. While independent gate operations are a key concept for quantum computing, it is actually not possible to fully describe noisy gates locally in time if noise is correlated on times longer than their duration. To address this issue, we develop a method based on the filter function formalism to perturbatively compute quantum processes in the presence of correlated classical noise. We derive a composition rule for the filter function of a sequence of gates in terms of those of the individual gates. The joint filter function allows to efficiently compute the quantum process of the whole sequence. Moreover, we show that correlation terms arise which capture the effects of the concatenation and thus yield insight into the effect of noise correlations on gate sequences. Our generalization of the filter function formalism enables both qualitative and quantitative studies of algorithms and state-of-the-art tools widely used for the experimental verification of gate fidelities like randomized benchmarking, even in the presence of noise correlations.
\end{abstract}

\maketitle

\emph{Introduction.}
A key concept in gate-based quantum computing is the composition of algorithms from a universal set of quantum gates. In real physical devices, gate implementations are subject to noise that causes decoherence and gate errors. If this noise is uncorrelated on timescales larger than the gate duration, each gate can still be described individually by a quantum operation acting on density matrices. A closely related approach is the use of a Master equation in Lindblad form \cite{Lindblad1976}, which governs the dynamics of density matrices under the influence of Markovian noise (defined here as noise that is uncorrelated on the time scale of the system dynamics). 

However, the assumption of uncorrelated noise is often unjustified. A prominent example is \oneoverf noise characteristic for flux noise in superconducting qubits and electrical noise in quantum dot qubits and ion trap qubits, which are among the most important types of noise for solid state qubits \cite{Brownnutt2015,Kumar2016, Yoneda2018,  Paladino2014}. Hence, the standard tools for mathematically describing gate operations are not suited for capturing experimentally relevant effects that are important for understanding the capabilities of quantum computing systems. Furthermore, the process description of a gate sequence can deviate from the concatenation of the individual gates' processes. For example, one may expect the fidelity requirements for quantum error correction to be more stringent for correlated noise as errors of different gates can interfere constructively \cite{Ng2009}.

Here, we present an intuitive and computationally efficient method based on the \gls{ff} formalism \cite{Biercuk2009,Soare2014,Malinowski2017,Paz-Silva2014} that overcomes these limitations for the purpose of computing process descriptions for arbitrary sequences of gate operations subject to correlated, classical Gaussian noise. This makes our approach attractive for studying the noise properties of quantum algorithms as we demonstrate with a simple example. Because widely used tools for the experimental verification of gate fidelities, such as \gls{gst} and \gls{rb}, rely on gate sequences, our approach can also shed light on the applicability of these protocols in the presence of correlated classical noise \cite{Ball2015,Mavadia2018,Edmunds2020} which violates a core assumption of standard derivations \cite{Magesan2011,Blume-Kohout2017}. Furthermore, our approach captures corresponding corrections in terms of \glspl{ff}.

\Glspl{ff} were originally introduced to compute the decay of phase coherence under dynamical decoupling sequences \cite{Kofman2001,Martinis2003,Cywinski2008,Uhrig2007} consisting of wait times and perfect $\pi$-pulses. For small noise strengths, they were perturbatively extended to quantum gates \cite{Green2012,Green2013,Gungordu2018,Ball2020} to compute gate fidelities \cite{Green2013,Gungordu2018,Vandijk2018} and develop strategies for noise mitigation \cite{Cerfontaine2014, Khodjasteh2013, Ball2015, Huang2017}. Here, we build on these results to compute quantum processes for gates and gate sequences on an arbitrary number of qubits and to analyse their concatenation properties. Relevant quantities like fidelities, measurement statistics, and leakage can be extracted from process descriptions or directly from corresponding filter functions. For ease of adoption, we also provide an easy-to-use Python software package \cite{Hangleiter2021,software}.

\emph{Quantum processes.}
We begin by deriving an approximate form of the average quantum process of a quantum gate of duration $\tau$ in the presence of arbitrary classical noise. Our approach builds on the fidelity calculations of \citerr{Green2012}{Green2013}, which we briefly review and generalize in some points. Concretely, we consider a system described by the Hamiltonian $H(t) = \Hc(t) + \Hn(t)$. The arbitrary, time-dependent control Hamiltonian $\Hc(t)$ generates the desired unitary evolution $\Uc(t)$. This evolution is perturbed by the noise Hamiltonian $\Hn(t) = \sum_\alpha b_\alpha(t) \Ba(t)$ which contains zero-mean, independent and identically distributed, classical Gaussian noise variables $b_\alpha(t)$. We generalize beyond ealier work by additionally allowing for a deterministic time dependence of the noise operators $\Ba(t)$ in \Hnt but for simplicity restrict ourselves to independent noise sources $\alpha$ and refer to \citer{Hangleiter2021} for the straightforward extension to cross-correlated noise.

Next, we write the propagator for $H(t)$ as $U(t) = \Uc(t)\Ue(t)$ where the unitary error propagator $\Ue(t)$ contains the effect of a specific noise realization. We transform \Hn to the interaction picture with respect to the control Hamiltonian, $\Hnt(t)\coloneqq\Uc^\dagger(t)\Hn(t)\Uc(t)$, so that $\Ue(t)$ satisfies $\i\:\dd\Ue(t)/\dd t = \Hnt(t)\Ue(t)$. Note that we set $\hbar = 1$ and denote operators in the interaction picture by a tilde throughout this work. $\Ue(t=\tau) \equiv \Ue$ can be generated by an effective Hamiltonian $H_\mr{eff}$ without explicit time dependence, $\Ue = \exp(-\i H_\mr{eff}\tau)$.

This effective Hamiltonian can be expanded using the \gls{me} $H_\mr{eff} = \sum_{\mu=1}^\infty H_\mr{eff,\mu}$. Since the \gls{me} preserves the algebraic structure of the expanded quantity, $H_\mr{eff}$ remains Hermitian even after truncating the series, allowing us to neglect contributions from higher orders. The first and second \gls{me} term are given by  $H_\mr{eff,1} = \flatfrac{1}{\tau}\int_0^\tau\dd{t}\Hnt(t)$ and $H_\mr{eff,2} = -\flatfrac{\i}{2\tau}\int_0^\tau\dd{t_1}\int_0^{t_1}\dd{t_2}\comm{\Hnt(t_1)}{\Hnt(t_2)}$, respectively \cite{Magnus1954,Blanes2009}. Higher orders provide diminishing contributions if the noise strength $\xi \coloneqq \sum_\alpha\norm{B_\alpha}\sigma_\alpha\tau\ll 1$, where $\sigma_\alpha = \expval*{b_\alpha(0)^2}^{1/2}$ is the standard deviation of the noise, because they include an increasing number of factors of \Hnt which we assumed to be small \footnote{While $\xi$ given here is only valid for time-independent $\Ba$, an extended discussion of the convergence criteria is given in \citer{Hangleiter2021} and \citer{Green2013}.}. This may be interpreted as the condition that the angle by which a specific noise realization $b_\alpha(t)$ has rotated the (generalized) Bloch vector away from its intended trajectory after time $\tau$ must be small.

We proceed beyond the works by \citet{Green2012}, where \Ue was used to compute the gate fidelity, to compute the full, noise-averaged quantum process $\Pe(\rho) \coloneqq \expval*{\Ue\rho\Ue^\dagger}$. We expand the error propagator \Ue in a Taylor series, keeping terms up to and including $\order{\xi^2}$, which yields (see also \citerr{Haeberlen1976}{Majenz2013}),
\begin{multline}\label{eq:ff_pe_all}
    \frac{\Pe(\rho) - \rho}{\tau} = -\i\langle\comm{H_\mr{eff,2}}{\rho}\rangle \\
                                    +\tau\langle H_\mr{eff,1}\rho H_\mr{eff,1} - \frac{1}{2}\acomm{H_\mr{eff,1}^2}{\rho}\rangle + \order{\xi^4},
\end{multline}
where square (curly) brackets denote the (anti-) commutator and $\langle \cdot \rangle$ represents averaging over noise realizations. We have already dropped terms that vanish after performing the average either due to $\expval{H_\mr{eff,1}} = 0$ or because correlation functions evaluated at an odd number of time points vanish for zero-mean Gaussian noise \cite{Blanes2009}. The form of \cref{eq:ff_pe_all} is reminiscent of a master equation in Lindblad form with Hamiltonian $H_\mr{eff,2}$ and jump operators $H_\mr{eff,1}$ with associated decay rate $\tau$. However, instead of a differential equation governing the time evolution of $\rho$, it represents a finite difference equation with $\dv*{\rho}{t}\rightarrow\flatfrac{\Delta\rho}{\tau}$ that describes the average evolution after the time $\tau$ at which the gate has completed. For a single qubit, $H_\mr{eff,2}$ generates a rotation, whereas the terms involving $H_\mr{eff,1}$ correspond to a deformation of the Bloch sphere into an ellipsoid.

While it is possible to evaluate both first and second order \gls{me} terms, it is more computationally involved to calculate the nested integrals contained in $H_\mr{eff,2}$ (see \citer{Green2013} for an explicit treatment of higher orders). However, we argue that these terms are of less interest in typical use cases. First, they vanish under the trace, and hence do not contribute to the fidelity of the quantum operation, $\fid\propto\tr\Pe$. Furthermore, second order \gls{me} terms represent the unitary (Hamiltonian) part of \cref{eq:ff_pe_all} that can be cancelled to leading order by a unitary rotation as commutators of $H_\mr{eff,1}$ and $H_\mr{eff,2}$ are $\order{\xi^{3}}$. Thus, it is possible to calculate $\Pe(\rho)$ up to a unitary rotation just by using first order \gls{me} terms. In many contexts, this is sufficient since unitary errors are typically calibrated out in experiments, using a variety of methods \cite{Cerfontaine2019gsc,Kimmel2015,Blume-Kohout2017,Reed2013,Kelly2014,Egger2014}. However, our work shows that even if all individual gates are perfectly calibrated, a sequence of gates might incur an additional unitary error not removed by individual calibration. Moreover, calibration procedures using gate sequences may be affected by such noise-induced coherent error. To study such effects, second order \gls{me} terms can be evaluated following a procedure we lay out in \citer{Hangleiter2021}. Finally, we note that truncating the expansion in \cref{eq:ff_pe_all} can in principle lead to unphysical dynamics, in the sense that the truncated map is not \gls{cp} \footnote{Note that this affects all perturbative approaches, \eg \citer{Green2013}}. In practice, this should not pose a relevant limitation because $\Pe(\rho)$ differs from the true final state by terms of $\order{\xi^4}$, leading to errors of the same order in measurement results. Thus, unphysical errors should be small as long as the perturbative expansion is well-defined. We have verified this hypothesis in (random) numerical experiments and found negative Choi eigenvalues \cite{Choi1975} to be comparatively small in magnitude, should they occur at all.

For Gaussian noise, it is possible to go beyond our perturbative treatment via an exact solution requiring only first and second order \gls{me} terms by applying the method of cumulant expansions to a stochastic Liouville equation \cite{Hangleiter2021}. As it turns out, this solution is simply the matrix exponential of the superoperator form of \cref{eq:ff_pe_all}, so that the latter takes on the role of the generator. Because \cref{eq:ff_pe_all} is in Lindblad form, it follows that its exponential is a \gls{cp} map \cite{Hall2014}. While the exact solution thus guarantees a physical output state, one loses qualitative insight into contributions from, for instance, different noise operators \Ba because the matrix exponential can in general only be evaluated numerically. Further details, including the evaluation of second order \gls{me} terms, are given in our related work \cite{Hangleiter2021}. Here, we focus on the non-unitary part of the weak-noise approximation \cref{eq:ff_pe_all} and now describe how to evaluate it.

We turn to the \gls{ff} formalism and express correlation functions of noise variables by their power spectral density and the evolution of the interaction picture noise operators by their \glspl{ff} in the Fourier domain. Expanding $\Hnt(t) = \sum_\alpha b_\alpha(t)\Bat(t)$ in a Hermitian and orthonormal operator basis $\lbrace\sigma_k\rbrace_{k=0}^{d^2-1}$ satisfying $\sigma_k\ad=\sigma_k$ and $\tr(\sigma_k\sigma_l)=\delta_{kl}$, we obtain
\begin{equation}\label{eq:ff_control_matrix}
    \Hnt(t) = \sum_{\alpha k} b_\alpha(t)\Rc_{\alpha k}(t)\sigma_k.
\end{equation}
A simple choice for the $\sigma_k$ is the $n$-qubit Pauli basis $\lbrace\eye,\px,\py,\pz\rbrace^{\otimes n}$ which we use in the following. We identify the coefficients of the expansion,
\begin{align}
\Rc_{\alpha k}(t) = \tr(\Bat(t)\sigma_k) = \tr(\Uc^\dagger(t) \Ba(t)\Uc(t)\sigma_k),
\end{align}
as the control matrix from \citer{Green2013} (which is related to the Pauli transfer matrix representation of a quantum process). Inserting \cref{eq:ff_control_matrix} into the effective master equation \cref{eq:ff_pe_all} and, dropping second order \gls{me} terms as justified above, we find
\begin{align}
    \Pe(\rho) - \rho &\approx \sum_{\alpha}\sum_{k l}\decayamps_{\alpha,kl}\left(\sigma_k\rho\sigma_l - \frac{1}{2}\acomm{\sigma_k\sigma_l}{\rho}\right) \label{eq:ff_pe_transformed}
\end{align}
with the matrix of decay amplitudes $\decayamps_\alpha$ with entries
\begin{equation}\label{eq:ff_decay_amplitudes:time_domain}
    \decayamps_{\alpha,kl} = \int_0^\tau\int_0^\tau\dd{t_1}\dd{t_2}\expval{b_\alpha(t_1) b_\alpha(t_2)}\Rc_{\alpha k}(t_1)\Rc_{\alpha l}(t_2).
\end{equation}
With this basis expansion, we have transformed the effective master equation to a basis in which the jump operators $\sigma_k$ are time-independent and only the decay amplitudes are functions of the internal dynamics of the gate and the noise. Hence, we can carry out the integration not on the operator level of the effective master equation but on the level of the decay amplitudes $\decayamps_{\alpha}$. This allows us to employ the \gls{ff} formalism to evaluate $\decayamps_{\alpha}$ in Fourier space. We define the two-sided noise spectral density $S_{\alpha}(\omega)$ as the Fourier transform of the autocorrelation function of the noise variable $b_\alpha(t)$ via
\begin{equation}\label{eq:ff_spectral_density}
    \expval{b_\alpha(t_1) b_{\alpha}(t_2)} = \int_{-\infty}^{\infty}\frac{\dd{\omega}}{2\pi} S_{\alpha}(\omega)\e^{-\i\omega (t_1 - t_2)},
\end{equation}
where we assume that the noise is wide-sense stationary. Inserting into \cref{eq:ff_decay_amplitudes:time_domain} yields
\begin{equation}\label{eq:ff_decay_amplitudes:fourier_domain}
    \decayamps_{\alpha,kl} = \int_{-\infty}^{\infty}\frac{\dd{\omega}}{2\pi} S_{\alpha}(\omega)\Rc_{\alpha k}^\ast(\omega)\Rc_{\alpha l}(\omega)
\end{equation} 
with $\Rc_{\alpha k}(\omega) = \int_0^\tau\dd{t} \Rc_{\alpha k}(t)\e^{\i\omega t}$. The generalized \gls{ff} $F_{\alpha, kl}(\omega) = \Rc_{\alpha k}^\ast(\omega)\Rc_{\alpha l}(\omega)$  describes the sensitivity of the decay amplitudes $\decayamps_{\alpha,kl}$ to noise source $\alpha$ at frequency $\omega$. We can now use \cref{eq:ff_pe_transformed} together with \cref{eq:ff_decay_amplitudes:fourier_domain} to obtain the quantum process $\Pe(\rho)$ generated by all noise sources up to first order \gls{me} and second order in $\xi$.

Given \Pe, it is straightforward to calculate key figures of merit for quantum gate operations like gate fidelity, leakage (i.e., the probability to leave the subspace of valid computational states of a physical system whose Hilbert space is often larger than the computational subspace), the diamond distance to the identity, or expectation values of measurements based on known relations, as we lay out in detail in \citer{Hangleiter2021} \footnote{See Supplemental Material \cite{prlSupp} for a brief overview, which includes Refs. \citenum{Wood2018,Wallman2016,Johansson2013,Schulte-herbruggen2005,Gorini1976}}. It is also possible to define specific \glspl{ff} that allow to directly compute these quantities from the spectral density. For example, consider the average gate fidelity to the identity \cite{Nielsen2002,Kimmel2014} given by $\fid = (\tr\Pe + d)/d(d+1)$ for whose evaluation only first order Magnus terms are relevant (\cf \cref{eq:ff_pe_all}). We obtain
\begin{equation}\label{eq:ff_fidelity}
    \fid = 1 - \frac{1}{d+1}\sum_{\alpha k}\decayamps_{\alpha, kk}
\end{equation}
where, in line with previous literature \cite{Green2012,Green2013}, we can identify the fidelity \gls{ff} up to first order \gls{me} as $F_{\alpha}(\omega) = \sum_k\lvert\Rc_{\alpha k}(\omega)\rvert^2$ which captures the fidelity's susceptibility to noise source $\alpha$ at frequency $\omega$.

\emph{Filter functions of gate sequences.}
We now show that the interaction picture noise operators \Bat for concatenated gates follow a simple composition rule that arises because subsequent gates update the frame of reference for the interaction picture. In the frequency domain, the total noise operators can be described as linear combinations of the single-gate noise operators, each multiplied with a phase factor corresponding to the gates' temporal positions. Since filter functions are quadratic in the noise operators, there arise correlation terms between \glspl{ff} at different positions in a gate sequence which constitute corrections to the \glspl{ff} of the separate gates. Our initial goal is to compute the decay amplitudes $\decayamps_\alpha$ for a sequence of quantum gates, and we will later on use these results to single out corrections arising from the concatenation alone. 

We consider that the control $\Uc(t)$ is implemented by concatenating several gates $P_g \equiv \Uc(t_g, t_{g-1}), ~ g\in\{1, 2,\dotsc, G\}$ with $t_0\equiv 0, t_G\equiv\tau$. Accordingly, we define the cumulative propagators $Q_g = P_g P_{g-1}\cdots P_0$ with $P_0\equiv\eye$ such that the total control operation is given by $Q\equiv Q_G$. Denoting by $\Qc\gth{g-1}(\bullet) = Q_{g-1}\ad\bullet Q_{g-1}$ the superoperator transforming to the interaction picture with respect to $Q_{g-1}$, we can write the interaction picture noise operators at time $t\in (t_{g-1}, t_g]$ as
\begin{equation}\label{eq:ff_control_matrix:time_domain}
    \Bat(t) = \Qc\gth{g-1}\left(\Bat\gth{g}(t - t_{g-1})\right),
\end{equation}
where $\Bat\gth{g}(t)$ are the noise operators in the interaction picture of the $g$th gate. We obtain the Fourier transform of $\Bat(t)$ by splitting up the integral into the time intervals $(t_{g-1}, t_g]$,
\begin{equation}\label{eq:ff_control_matrix:fourier_domain}
    \Bat(\omega)  = \sum_{g=1}^G \e^{\i\omega t_{g-1}}\Qc\gth{g-1}\left(\Bat\gth{g}(\omega)\right),
\end{equation}
with $\Bat\gth{g}(\omega) = \int_0^{\Delta t_g}\dd{t}\e^{\i\omega t}\Bat\gth{g}(t)$ and $\Delta t_g = t_g - t_{g-1}$ (\cf \citer{Green2013}). The terms $\Rc_{\alpha k}(\omega) = \mr{tr}\bigl(\Bat(\omega)\sigma_k\bigr)$ can now be used to calculate the generalized \glspl{ff} $F_{\alpha, kl}(\omega)$. \Cref{eq:ff_control_matrix:fourier_domain} thus illustrates that generalized \glspl{ff} of an entire sequence of gates can be easily calculated if the noise operators $\Bat\gth{g}(\omega)$ of the single gates have already been computed, for example by following \citer{Green2012} or \citer{Hangleiter2021}.

\emph{Correlation filter functions.}
By combining \cref{eq:ff_pe_transformed,eq:ff_control_matrix:fourier_domain}, we find leading-order corrections to the quantum process of a sequence of gates that arise solely from the concatenation operation itself and hence allow valuable insight into effects relevant for algorithms. We call these corrections, which depend on the positions $(g, g')$ of two gates in a sequence with $g=g'$ corresponding to the regular \gls{ff} of the $g$th gate, \glspl{cff}. We explicitly show this relation for the well-known fidelity \gls{ff}, but as with regular \glspl{ff}, one may also derive \glspl{cff} for other quantities as linear combinations of generalized \glspl{cff}. We use \cref{eq:ff_fidelity} together with \cref{eq:ff_decay_amplitudes:fourier_domain,eq:ff_control_matrix:fourier_domain} to compute the infidelity $\infid = 1 - \fid$ and find that
\begin{equation}\label{eq:ff_main_infidelity:concatenated}
    \infid \eqqcolon \sum_{g=1}^{G}\biggl[\infid\gth{g} + \frac{1}{d+1}\sum_\alpha\sum_{\substack{g'=1\\g'\neq g}}^{G}
                     \int_{-\infty}^{\infty}\frac{\dd{\omega}}{2\pi} S_\alpha(\omega)F_\alpha^{(gg')}(\omega)
    \biggr],
\end{equation}
where $\infid\gth{g}$ is the infidelity of the $g$th pulse alone and $F_\alpha^{(gg')}(\omega)$ \footnote{$F_\alpha^{(gg')}(\omega) = \e^{\i\omega(t_{g-1} - t_{g'-1})}\times\\\mr{tr}\bigl(\Qc\gth{g'-1}\bigl(\Bat\gth{g'}(\omega)\bigr)\ad\Qc\gth{g-1}\bigl(\Bat\gth{g}(\omega)\bigr)\bigr)$} is a \gls{ff} describing correlation effects between the pulses at positions $g$ and $g'$ in the sequence due to noise source $\alpha$. By summing over all gates the regular fidelity \gls{ff} $F_\alpha(\omega) = \sum_{g,g'=1}^{G} F_\alpha^{(gg')}(\omega)$ can be obtained. Unlike $F_\alpha(\omega)$, $F_\alpha^{(gg')}(\omega)$ is complex-valued and not strictly positive (but Hermitian in $g$ and $g'$ so that the sum over all $g,g'$ is real). Moreover, since second order \gls{me} terms are traceless, \cref{eq:ff_main_infidelity:concatenated} is exact up to $\order{\xi^4}$.

\glspl{cff} can, for example, capture the effect of dynamical error suppression in \gls{se} experiments, which we can view as a sequence of an idle pulse, a $\pi$ pulse, and another idle pulse. The regular \glspl{ff} of each of these individual pulses, that is $F_\alpha^{(gg)}(\omega)$ for $g\in\lbrace 1, 2, 3\rbrace$, are characterized by a finite value at low frequencies as the idle pulses simply correspond to free induction decays with \gls{ff} $\simeq\flatfrac{\sin^2(\flatfrac{\omega t_g}{2})}{\omega^2}$ \cite{Cywinski2008}. The error cancellation for low frequency dephasing ($\sigma_z$) noise then arises mainly from the \gls{cff} $F_z^{(1,3)}(\omega)\simeq -\flatfrac{\sin^2(\flatfrac{\omega \tau_\mr{idle}}{2})\exp(\i\omega \tau_\mr{idle})}{\omega^2}$ between the two idle pulses which we may interpret as destructive interference.
\begin{figure}
    \centering
    \includegraphics{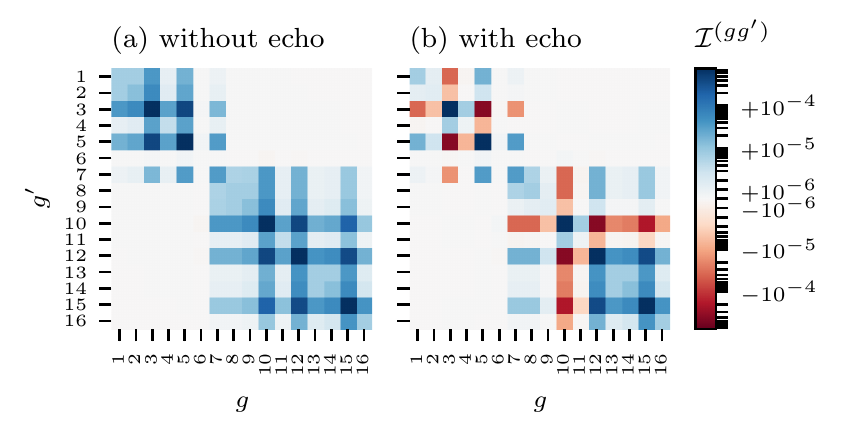}
    \caption{Correlation infidelities $\mc{I}\gth{gg'}$ for a four-qubit \gls{qft} circuit without (a) and with (b) additional $\pi_x$-pulses on the fourth qubit. We only consider \py-noise on the fourth qubit modelled by a $\flatfrac{1}{f}$ spectral noise density. The indices $g,g'=1,\dotsc,16$ indicate the gates' temporal positions in the circuit. Without the echo pulses, correlations in the noise give rise to significant infidelity contributions for gate pairs up to ten clock cycles apart. Incorporating echoes leads to negative correlation infidelities between pulses, which in turn decreases the total infidelity, given by $\sum_{gg'}\mc{I}\gth{gg'}$, by more than a factor 3 for this noise channel.}
    \label{fig:correlation_infidelities}
\end{figure}
As a more involved example, consider a quantum Fourier transform \cite{Coppersmith1994} on four qubits coupled via nearest neighbor interactions \footnote{Example code available at \url{https://github.com/qutech/filter_functions/blob/master/doc/source/examples/quantum_fourier_transform.ipynb}}. Using our approach, we investigate the effects of adding spin echos to idling qubits on the total algorithm's fidelity. We apply four $\pi_x$-pulses on the fourth qubit, two before a controlled phase gate and two afterwards \footnote{See Supplemental Material \cite{prlSupp} for the circuit diagram and the physical model}, and compute the correlation infidelities $\infid\gth{gg'} = \frac{1}{d+1}\sum_\alpha\int_{-\infty}^{\infty}\frac{\dd{\omega}}{2\pi} S_\alpha(\omega)F_\alpha^{(gg')}(\omega)$ between different pulses (the second term in \cref{eq:ff_main_infidelity:concatenated}). The results are shown in \cref{fig:correlation_infidelities} for \oneoverf-noise on \py on the fourth qubit. Certain pairs of gates have negative correlation infidelities on the order of magnitude of individual gate infidelities, indicating that they cancel errors to a large degree. Indeed, a more than threefold reduction of the total infidelity is observed. Repeating the analysis for white noise reveals that the echo pulses do not change the fidelity significantly.

\emph{Conclusion and outlook.}
In this work we have shown how to efficiently obtain process matrices on an arbitrary number of qubits in the presence of correlated classical noise. Many relevant quantities can easily be derived from such a process description, including fidelity measures, measurement statistics, and leakage. In addition, we have introduced the concept of \glspl{cff}, which describe corrections when sequences of gates are executed in the presence of noise correlations. As such, \glspl{cff} are particularly relevant for testing the notion of independent gates in quantum computing applications, and can be used to calculate correction terms when this is not the case. \glspl{cff} also facilitate the analysis of larger circuits by recycling \glspl{ff} already computed for individual gates.

We also provide a user-friendly and computationally efficient open source software package \cite{software}. This package, an extension to arbitrary bases, the calculation of several derived quantities, computational efficiency improvements including periodic driving Hamiltonians and exact results for Gaussian noise based on the cumulant expansion are described in greater detail in \citer{Hangleiter2021}. The latter is particularly relevant for dynamically corrected gates that decouple to lowest order from correlated noise so that second order terms can become dominant \cite{Green2012}.

We expect our approach to be useful for analyzing and improving the performance of experimental systems comprising several qubits, for example by leveraging optimal control approaches \cite{Le2021}. \glspl{ff} can be more efficient than Monte Carlo methods and directly allow further insight into qualitative effects of different types of noise spectra. Possible applications include the analysis and construction of novel dynamical decoupling sequences, noise spectroscopy protocols, dynamically corrected gates, small algorithms and \gls{qec} protocols. Our results could also facilitate the development of more realistic qubit benchmarking protocols, which fully take noise correlations into account.

\emph{Acknowledgements.}
We thank Maarten Wegewijs for helpful comments on the manuscript.
This work was supported by the European Research Council (ERC) under the European Union's Horizon 2020 research and innovation program (Grant Agreement No. 679342). P. C. and T. H. contributed equally to this work.


\begin{thebibliography}{56}%
\makeatletter
\providecommand \@ifxundefined [1]{%
 \@ifx{#1\undefined}
}%
\providecommand \@ifnum [1]{%
 \ifnum #1\expandafter \@firstoftwo
 \else \expandafter \@secondoftwo
 \fi
}%
\providecommand \@ifx [1]{%
 \ifx #1\expandafter \@firstoftwo
 \else \expandafter \@secondoftwo
 \fi
}%
\providecommand \natexlab [1]{#1}%
\providecommand \enquote  [1]{``#1''}%
\providecommand \bibnamefont  [1]{#1}%
\providecommand \bibfnamefont [1]{#1}%
\providecommand \citenamefont [1]{#1}%
\providecommand \href@noop [0]{\@secondoftwo}%
\providecommand \href [0]{\begingroup \@sanitize@url \@href}%
\providecommand \@href[1]{\@@startlink{#1}\@@href}%
\providecommand \@@href[1]{\endgroup#1\@@endlink}%
\providecommand \@sanitize@url [0]{\catcode `\\12\catcode `\$12\catcode
  `\&12\catcode `\#12\catcode `\^12\catcode `\_12\catcode `\%12\relax}%
\providecommand \@@startlink[1]{}%
\providecommand \@@endlink[0]{}%
\providecommand \url  [0]{\begingroup\@sanitize@url \@url }%
\providecommand \@url [1]{\endgroup\@href {#1}{\urlprefix }}%
\providecommand \urlprefix  [0]{URL }%
\providecommand \Eprint [0]{\href }%
\providecommand \doibase [0]{http://dx.doi.org/}%
\providecommand \selectlanguage [0]{\@gobble}%
\providecommand \bibinfo  [0]{\@secondoftwo}%
\providecommand \bibfield  [0]{\@secondoftwo}%
\providecommand \translation [1]{[#1]}%
\providecommand \BibitemOpen [0]{}%
\providecommand \bibitemStop [0]{}%
\providecommand \bibitemNoStop [0]{.\EOS\space}%
\providecommand \EOS [0]{\spacefactor3000\relax}%
\providecommand \BibitemShut  [1]{\csname bibitem#1\endcsname}%
\let\auto@bib@innerbib\@empty
%</preamble>
\bibitem [{\citenamefont {Lindblad}(1976)}]{Lindblad1976}%
  \BibitemOpen
  \bibfield  {author} {\bibinfo {author} {\bibfnamefont {G.}~\bibnamefont
  {Lindblad}},\ }\href@noop {} {\bibfield  {journal} {\bibinfo  {journal}
  {Communications in Mathematical Physics}\ }\textbf {\bibinfo {volume} {48}},\
  \bibinfo {pages} {119} (\bibinfo {year} {1976})}\BibitemShut {NoStop}%
\bibitem [{\citenamefont {Brownnutt}\ \emph {et~al.}(2015)\citenamefont
  {Brownnutt}, \citenamefont {Kumph}, \citenamefont {Rabl},\ and\ \citenamefont
  {Blatt}}]{Brownnutt2015}%
  \BibitemOpen
  \bibfield  {author} {\bibinfo {author} {\bibfnamefont {M.}~\bibnamefont
  {Brownnutt}}, \bibinfo {author} {\bibfnamefont {M.}~\bibnamefont {Kumph}},
  \bibinfo {author} {\bibfnamefont {P.}~\bibnamefont {Rabl}}, \ and\ \bibinfo
  {author} {\bibfnamefont {R.}~\bibnamefont {Blatt}},\ }\href {\doibase
  10.1103/RevModPhys.87.1419} {\bibfield  {journal} {\bibinfo  {journal} {Rev.
  Mod. Phys.}\ }\textbf {\bibinfo {volume} {87}},\ \bibinfo {pages} {1419}
  (\bibinfo {year} {2015})}\BibitemShut {NoStop}%
\bibitem [{\citenamefont {Kumar}\ \emph {et~al.}(2016)\citenamefont {Kumar},
  \citenamefont {Sendelbach}, \citenamefont {Beck}, \citenamefont {Freeland},
  \citenamefont {Wang}, \citenamefont {Wang}, \citenamefont {Yu}, \citenamefont
  {Wu}, \citenamefont {Pappas},\ and\ \citenamefont {McDermott}}]{Kumar2016}%
  \BibitemOpen
  \bibfield  {author} {\bibinfo {author} {\bibfnamefont {P.}~\bibnamefont
  {Kumar}}, \bibinfo {author} {\bibfnamefont {S.}~\bibnamefont {Sendelbach}},
  \bibinfo {author} {\bibfnamefont {M.~A.}\ \bibnamefont {Beck}}, \bibinfo
  {author} {\bibfnamefont {J.~W.}\ \bibnamefont {Freeland}}, \bibinfo {author}
  {\bibfnamefont {Z.}~\bibnamefont {Wang}}, \bibinfo {author} {\bibfnamefont
  {H.}~\bibnamefont {Wang}}, \bibinfo {author} {\bibfnamefont {C.~C.}\
  \bibnamefont {Yu}}, \bibinfo {author} {\bibfnamefont {R.~Q.}\ \bibnamefont
  {Wu}}, \bibinfo {author} {\bibfnamefont {D.~P.}\ \bibnamefont {Pappas}}, \
  and\ \bibinfo {author} {\bibfnamefont {R.}~\bibnamefont {McDermott}},\ }\href
  {\doibase 10.1103/PhysRevApplied.6.041001} {\bibfield  {journal} {\bibinfo
  {journal} {Phys. Rev. Applied}\ }\textbf {\bibinfo {volume} {6}},\ \bibinfo
  {pages} {041001(R)} (\bibinfo {year} {2016})}\BibitemShut {NoStop}%
\bibitem [{\citenamefont {Yoneda}\ \emph {et~al.}(2018)\citenamefont {Yoneda},
  \citenamefont {Takeda}, \citenamefont {Otsuka}, \citenamefont {Nakajima},
  \citenamefont {Delbecq}, \citenamefont {Allison}, \citenamefont {Honda},
  \citenamefont {Kodera}, \citenamefont {Oda}, \citenamefont {Hoshi},
  \citenamefont {Usami}, \citenamefont {Itoh},\ and\ \citenamefont
  {Tarucha}}]{Yoneda2018}%
  \BibitemOpen
  \bibfield  {author} {\bibinfo {author} {\bibfnamefont {J.}~\bibnamefont
  {Yoneda}}, \bibinfo {author} {\bibfnamefont {K.}~\bibnamefont {Takeda}},
  \bibinfo {author} {\bibfnamefont {T.}~\bibnamefont {Otsuka}}, \bibinfo
  {author} {\bibfnamefont {T.}~\bibnamefont {Nakajima}}, \bibinfo {author}
  {\bibfnamefont {M.~R.}\ \bibnamefont {Delbecq}}, \bibinfo {author}
  {\bibfnamefont {G.}~\bibnamefont {Allison}}, \bibinfo {author} {\bibfnamefont
  {T.}~\bibnamefont {Honda}}, \bibinfo {author} {\bibfnamefont
  {T.}~\bibnamefont {Kodera}}, \bibinfo {author} {\bibfnamefont
  {S.}~\bibnamefont {Oda}}, \bibinfo {author} {\bibfnamefont {Y.}~\bibnamefont
  {Hoshi}}, \bibinfo {author} {\bibfnamefont {N.}~\bibnamefont {Usami}},
  \bibinfo {author} {\bibfnamefont {K.~M.}\ \bibnamefont {Itoh}}, \ and\
  \bibinfo {author} {\bibfnamefont {S.}~\bibnamefont {Tarucha}},\ }\href
  {\doibase 10.1038/s41565-017-0014-x} {\bibfield  {journal} {\bibinfo
  {journal} {Nature Nanotechnology}\ }\textbf {\bibinfo {volume} {13}},\
  \bibinfo {pages} {102} (\bibinfo {year} {2018})}\BibitemShut {NoStop}%
\bibitem [{\citenamefont {Paladino}\ \emph {et~al.}(2014)\citenamefont
  {Paladino}, \citenamefont {Galperin}, \citenamefont {Falci},\ and\
  \citenamefont {Altshuler}}]{Paladino2014}%
  \BibitemOpen
  \bibfield  {author} {\bibinfo {author} {\bibfnamefont {E.}~\bibnamefont
  {Paladino}}, \bibinfo {author} {\bibfnamefont {Y.~M.}\ \bibnamefont
  {Galperin}}, \bibinfo {author} {\bibfnamefont {G.}~\bibnamefont {Falci}}, \
  and\ \bibinfo {author} {\bibfnamefont {B.~L.}\ \bibnamefont {Altshuler}},\
  }\href {\doibase 10.1103/RevModPhys.86.361} {\bibfield  {journal} {\bibinfo
  {journal} {Reviews of Modern Physics}\ }\textbf {\bibinfo {volume} {86}},\
  \bibinfo {pages} {361} (\bibinfo {year} {2014})}\BibitemShut {NoStop}%
\bibitem [{\citenamefont {Ng}\ and\ \citenamefont {Preskill}(2009)}]{Ng2009}%
  \BibitemOpen
  \bibfield  {author} {\bibinfo {author} {\bibfnamefont {H.~K.}\ \bibnamefont
  {Ng}}\ and\ \bibinfo {author} {\bibfnamefont {J.}~\bibnamefont {Preskill}},\
  }\href {\doibase 10.1103/PhysRevA.79.032318} {\bibfield  {journal} {\bibinfo
  {journal} {Phys. Rev. A}\ }\textbf {\bibinfo {volume} {79}},\ \bibinfo
  {pages} {032318} (\bibinfo {year} {2009})}\BibitemShut {NoStop}%
\bibitem [{\citenamefont {Biercuk}\ \emph {et~al.}(2009)\citenamefont
  {Biercuk}, \citenamefont {Uys}, \citenamefont {VanDevender}, \citenamefont
  {Shiga}, \citenamefont {Itano},\ and\ \citenamefont
  {Bollinger}}]{Biercuk2009}%
  \BibitemOpen
  \bibfield  {author} {\bibinfo {author} {\bibfnamefont {M.~J.}\ \bibnamefont
  {Biercuk}}, \bibinfo {author} {\bibfnamefont {H.}~\bibnamefont {Uys}},
  \bibinfo {author} {\bibfnamefont {A.~P.}\ \bibnamefont {VanDevender}},
  \bibinfo {author} {\bibfnamefont {N.}~\bibnamefont {Shiga}}, \bibinfo
  {author} {\bibfnamefont {W.~M.}\ \bibnamefont {Itano}}, \ and\ \bibinfo
  {author} {\bibfnamefont {J.~J.}\ \bibnamefont {Bollinger}},\ }\href {\doibase
  10.1038/nature07951} {\bibfield  {journal} {\bibinfo  {journal} {Nature}\
  }\textbf {\bibinfo {volume} {458}},\ \bibinfo {pages} {996} (\bibinfo {year}
  {2009})}\BibitemShut {NoStop}%
\bibitem [{\citenamefont {Soare}\ \emph {et~al.}(2014)\citenamefont {Soare},
  \citenamefont {Ball}, \citenamefont {Hayes}, \citenamefont {Sastrawan},
  \citenamefont {Jarratt}, \citenamefont {Mcloughlin}, \citenamefont {Zhen},
  \citenamefont {Green},\ and\ \citenamefont {Biercuk}}]{Soare2014}%
  \BibitemOpen
  \bibfield  {author} {\bibinfo {author} {\bibfnamefont {A.}~\bibnamefont
  {Soare}}, \bibinfo {author} {\bibfnamefont {H.}~\bibnamefont {Ball}},
  \bibinfo {author} {\bibfnamefont {D.}~\bibnamefont {Hayes}}, \bibinfo
  {author} {\bibfnamefont {J.}~\bibnamefont {Sastrawan}}, \bibinfo {author}
  {\bibfnamefont {M.~C.}\ \bibnamefont {Jarratt}}, \bibinfo {author}
  {\bibfnamefont {J.~J.}\ \bibnamefont {Mcloughlin}}, \bibinfo {author}
  {\bibfnamefont {X.}~\bibnamefont {Zhen}}, \bibinfo {author} {\bibfnamefont
  {T.~J.}\ \bibnamefont {Green}}, \ and\ \bibinfo {author} {\bibfnamefont
  {M.~J.}\ \bibnamefont {Biercuk}},\ }\href {\doibase 10.1038/nphys3115}
  {\bibfield  {journal} {\bibinfo  {journal} {Nature Physics}\ }\textbf
  {\bibinfo {volume} {10}},\ \bibinfo {pages} {825} (\bibinfo {year}
  {2014})}\BibitemShut {NoStop}%
\bibitem [{\citenamefont {Malinowski}\ \emph {et~al.}(2017)\citenamefont
  {Malinowski}, \citenamefont {Martins}, \citenamefont {Nissen}, \citenamefont
  {Barnes}, \citenamefont {Cywi{\'{n}}ski}, \citenamefont {Rudner},
  \citenamefont {Fallahi}, \citenamefont {Gardner}, \citenamefont {Manfra},
  \citenamefont {Marcus},\ and\ \citenamefont {Kuemmeth}}]{Malinowski2017}%
  \BibitemOpen
  \bibfield  {author} {\bibinfo {author} {\bibfnamefont {F.~K.}\ \bibnamefont
  {Malinowski}}, \bibinfo {author} {\bibfnamefont {F.}~\bibnamefont {Martins}},
  \bibinfo {author} {\bibfnamefont {P.~D.}\ \bibnamefont {Nissen}}, \bibinfo
  {author} {\bibfnamefont {E.}~\bibnamefont {Barnes}}, \bibinfo {author}
  {\bibfnamefont {{\L}.}~\bibnamefont {Cywi{\'{n}}ski}}, \bibinfo {author}
  {\bibfnamefont {M.~S.}\ \bibnamefont {Rudner}}, \bibinfo {author}
  {\bibfnamefont {S.}~\bibnamefont {Fallahi}}, \bibinfo {author} {\bibfnamefont
  {G.~C.}\ \bibnamefont {Gardner}}, \bibinfo {author} {\bibfnamefont {M.~J.}\
  \bibnamefont {Manfra}}, \bibinfo {author} {\bibfnamefont {C.~M.}\
  \bibnamefont {Marcus}}, \ and\ \bibinfo {author} {\bibfnamefont
  {F.}~\bibnamefont {Kuemmeth}},\ }\href {\doibase 10.1038/nnano.2016.170}
  {\bibfield  {journal} {\bibinfo  {journal} {Nature Nanotechnology}\ }\textbf
  {\bibinfo {volume} {12}},\ \bibinfo {pages} {16} (\bibinfo {year}
  {2017})}\BibitemShut {NoStop}%
\bibitem [{\citenamefont {Paz-Silva}\ and\ \citenamefont
  {Viola}(2014)}]{Paz-Silva2014}%
  \BibitemOpen
  \bibfield  {author} {\bibinfo {author} {\bibfnamefont {G.~A.}\ \bibnamefont
  {Paz-Silva}}\ and\ \bibinfo {author} {\bibfnamefont {L.}~\bibnamefont
  {Viola}},\ }\href {\doibase 10.1103/PhysRevLett.113.250501} {\bibfield
  {journal} {\bibinfo  {journal} {Phys. Rev. Lett.}\ }\textbf {\bibinfo
  {volume} {113}},\ \bibinfo {pages} {250501} (\bibinfo {year}
  {2014})}\BibitemShut {NoStop}%
\bibitem [{\citenamefont {Ball}\ and\ \citenamefont
  {Biercuk}(2015)}]{Ball2015}%
  \BibitemOpen
  \bibfield  {author} {\bibinfo {author} {\bibfnamefont {H.}~\bibnamefont
  {Ball}}\ and\ \bibinfo {author} {\bibfnamefont {M.~J.}\ \bibnamefont
  {Biercuk}},\ }\href {https://dx.doi.org/10.1140/epjqt/s40507-015-0022-4}
  {\bibfield  {journal} {\bibinfo  {journal} {Eur. Phys. J. Quantum Technol.}\
  }\textbf {\bibinfo {volume} {2}} (\bibinfo {year} {2015})}\BibitemShut
  {NoStop}%
\bibitem [{\citenamefont {Mavadia}\ \emph {et~al.}(2018)\citenamefont
  {Mavadia}, \citenamefont {Edmunds}, \citenamefont {Hempel}, \citenamefont
  {Ball}, \citenamefont {Roy}, \citenamefont {Stace},\ and\ \citenamefont
  {Biercuk}}]{Mavadia2018}%
  \BibitemOpen
  \bibfield  {author} {\bibinfo {author} {\bibfnamefont {S.}~\bibnamefont
  {Mavadia}}, \bibinfo {author} {\bibfnamefont {C.~L.}\ \bibnamefont
  {Edmunds}}, \bibinfo {author} {\bibfnamefont {C.}~\bibnamefont {Hempel}},
  \bibinfo {author} {\bibfnamefont {H.}~\bibnamefont {Ball}}, \bibinfo {author}
  {\bibfnamefont {F.}~\bibnamefont {Roy}}, \bibinfo {author} {\bibfnamefont
  {T.~M.}\ \bibnamefont {Stace}}, \ and\ \bibinfo {author} {\bibfnamefont
  {M.~J.}\ \bibnamefont {Biercuk}},\ }\href {\doibase
  10.1038/s41534-017-0052-0} {\bibfield  {journal} {\bibinfo  {journal} {npj
  Quantum Inf.}\ }\textbf {\bibinfo {volume} {4}},\ \bibinfo {pages} {7}
  (\bibinfo {year} {2018})}\BibitemShut {NoStop}%
\bibitem [{\citenamefont {Edmunds}\ \emph {et~al.}(2020)\citenamefont
  {Edmunds}, \citenamefont {Hempel}, \citenamefont {Harris}, \citenamefont
  {Frey}, \citenamefont {Stace},\ and\ \citenamefont {Biercuk}}]{Edmunds2020}%
  \BibitemOpen
  \bibfield  {author} {\bibinfo {author} {\bibfnamefont {C.~L.}\ \bibnamefont
  {Edmunds}}, \bibinfo {author} {\bibfnamefont {C.}~\bibnamefont {Hempel}},
  \bibinfo {author} {\bibfnamefont {R.~J.}\ \bibnamefont {Harris}}, \bibinfo
  {author} {\bibfnamefont {V.}~\bibnamefont {Frey}}, \bibinfo {author}
  {\bibfnamefont {T.~M.}\ \bibnamefont {Stace}}, \ and\ \bibinfo {author}
  {\bibfnamefont {M.~J.}\ \bibnamefont {Biercuk}},\ }\href {\doibase
  10.1103/PhysRevResearch.2.013156} {\bibfield  {journal} {\bibinfo  {journal}
  {Phys. Rev. Research}\ }\textbf {\bibinfo {volume} {2}},\ \bibinfo {pages}
  {013156} (\bibinfo {year} {2020})}\BibitemShut {NoStop}%
\bibitem [{\citenamefont {Magesan}\ \emph {et~al.}(2011)\citenamefont
  {Magesan}, \citenamefont {Gambetta},\ and\ \citenamefont
  {Emerson}}]{Magesan2011}%
  \BibitemOpen
  \bibfield  {author} {\bibinfo {author} {\bibfnamefont {E.}~\bibnamefont
  {Magesan}}, \bibinfo {author} {\bibfnamefont {J.~M.}\ \bibnamefont
  {Gambetta}}, \ and\ \bibinfo {author} {\bibfnamefont {J.}~\bibnamefont
  {Emerson}},\ }\href {\doibase 10.1103/PhysRevLett.106.180504} {\bibfield
  {journal} {\bibinfo  {journal} {Phys. Rev. Lett.}\ }\textbf {\bibinfo
  {volume} {106}},\ \bibinfo {pages} {180504} (\bibinfo {year}
  {2011})}\BibitemShut {NoStop}%
\bibitem [{\citenamefont {Blume-Kohout}\ \emph {et~al.}(2017)\citenamefont
  {Blume-Kohout}, \citenamefont {Gamble}, \citenamefont {Nielsen},
  \citenamefont {Rudinger}, \citenamefont {Mizrahi}, \citenamefont {Fortier},\
  and\ \citenamefont {Maunz}}]{Blume-Kohout2017}%
  \BibitemOpen
  \bibfield  {author} {\bibinfo {author} {\bibfnamefont {R.}~\bibnamefont
  {Blume-Kohout}}, \bibinfo {author} {\bibfnamefont {J.~K.}\ \bibnamefont
  {Gamble}}, \bibinfo {author} {\bibfnamefont {E.}~\bibnamefont {Nielsen}},
  \bibinfo {author} {\bibfnamefont {K.}~\bibnamefont {Rudinger}}, \bibinfo
  {author} {\bibfnamefont {J.}~\bibnamefont {Mizrahi}}, \bibinfo {author}
  {\bibfnamefont {K.}~\bibnamefont {Fortier}}, \ and\ \bibinfo {author}
  {\bibfnamefont {P.}~\bibnamefont {Maunz}},\ }\href
  {https://www.nature.com/doifinder/10.1038/ncomms14485} {\bibfield  {journal}
  {\bibinfo  {journal} {Nature Communications}\ }\textbf {\bibinfo {volume}
  {8}},\ \bibinfo {pages} {14485} (\bibinfo {year} {2017})}\BibitemShut
  {NoStop}%
\bibitem [{\citenamefont {Kofman}\ and\ \citenamefont
  {Kurizki}(2001)}]{Kofman2001}%
  \BibitemOpen
  \bibfield  {author} {\bibinfo {author} {\bibfnamefont {A.~G.}\ \bibnamefont
  {Kofman}}\ and\ \bibinfo {author} {\bibfnamefont {G.}~\bibnamefont
  {Kurizki}},\ }\href {\doibase 10.1103/PhysRevLett.87.270405} {\bibfield
  {journal} {\bibinfo  {journal} {Phys. Rev. Lett.}\ }\textbf {\bibinfo
  {volume} {87}},\ \bibinfo {pages} {270405} (\bibinfo {year}
  {2001})}\BibitemShut {NoStop}%
\bibitem [{\citenamefont {Martinis}\ \emph {et~al.}(2003)\citenamefont
  {Martinis}, \citenamefont {Nam}, \citenamefont {Aumentado}, \citenamefont
  {Lang},\ and\ \citenamefont {Urbina}}]{Martinis2003}%
  \BibitemOpen
  \bibfield  {author} {\bibinfo {author} {\bibfnamefont {J.~M.}\ \bibnamefont
  {Martinis}}, \bibinfo {author} {\bibfnamefont {S.}~\bibnamefont {Nam}},
  \bibinfo {author} {\bibfnamefont {J.}~\bibnamefont {Aumentado}}, \bibinfo
  {author} {\bibfnamefont {K.~M.}\ \bibnamefont {Lang}}, \ and\ \bibinfo
  {author} {\bibfnamefont {C.}~\bibnamefont {Urbina}},\ }\href {\doibase
  10.1103/PhysRevB.67.094510} {\bibfield  {journal} {\bibinfo  {journal} {Phys.
  Rev. B}\ }\textbf {\bibinfo {volume} {67}},\ \bibinfo {pages} {094510}
  (\bibinfo {year} {2003})}\BibitemShut {NoStop}%
\bibitem [{\citenamefont {Cywi{\'{n}}ski}\ \emph {et~al.}(2008)\citenamefont
  {Cywi{\'{n}}ski}, \citenamefont {Lutchyn}, \citenamefont {Nave},\ and\
  \citenamefont {{Das Sarma}}}]{Cywinski2008}%
  \BibitemOpen
  \bibfield  {author} {\bibinfo {author} {\bibfnamefont {{\L}.}~\bibnamefont
  {Cywi{\'{n}}ski}}, \bibinfo {author} {\bibfnamefont {R.~M.}\ \bibnamefont
  {Lutchyn}}, \bibinfo {author} {\bibfnamefont {C.~P.}\ \bibnamefont {Nave}}, \
  and\ \bibinfo {author} {\bibfnamefont {S.}~\bibnamefont {{Das Sarma}}},\
  }\href {\doibase 10.1103/PhysRevB.77.174509} {\bibfield  {journal} {\bibinfo
  {journal} {Phys. Rev. B}\ }\textbf {\bibinfo {volume} {77}},\ \bibinfo
  {pages} {174509} (\bibinfo {year} {2008})}\BibitemShut {NoStop}%
\bibitem [{\citenamefont {Uhrig}(2007)}]{Uhrig2007}%
  \BibitemOpen
  \bibfield  {author} {\bibinfo {author} {\bibfnamefont {G.~S.}\ \bibnamefont
  {Uhrig}},\ }\href {\doibase 10.1103/PhysRevLett.98.100504} {\bibfield
  {journal} {\bibinfo  {journal} {Phys. Rev. Lett.}\ }\textbf {\bibinfo
  {volume} {98}},\ \bibinfo {pages} {100504} (\bibinfo {year}
  {2007})}\BibitemShut {NoStop}%
\bibitem [{\citenamefont {Green}\ \emph {et~al.}(2012)\citenamefont {Green},
  \citenamefont {Uys},\ and\ \citenamefont {Biercuk}}]{Green2012}%
  \BibitemOpen
  \bibfield  {author} {\bibinfo {author} {\bibfnamefont {T.}~\bibnamefont
  {Green}}, \bibinfo {author} {\bibfnamefont {H.}~\bibnamefont {Uys}}, \ and\
  \bibinfo {author} {\bibfnamefont {M.~J.}\ \bibnamefont {Biercuk}},\ }\href
  {\doibase 10.1103/PhysRevLett.109.020501} {\bibfield  {journal} {\bibinfo
  {journal} {Phys. Rev. Lett.}\ }\textbf {\bibinfo {volume} {109}},\ \bibinfo
  {pages} {020501} (\bibinfo {year} {2012})}\BibitemShut {NoStop}%
\bibitem [{\citenamefont {Green}\ \emph {et~al.}(2013)\citenamefont {Green},
  \citenamefont {Sastrawan}, \citenamefont {Uys},\ and\ \citenamefont
  {Biercuk}}]{Green2013}%
  \BibitemOpen
  \bibfield  {author} {\bibinfo {author} {\bibfnamefont {T.~J.}\ \bibnamefont
  {Green}}, \bibinfo {author} {\bibfnamefont {J.}~\bibnamefont {Sastrawan}},
  \bibinfo {author} {\bibfnamefont {H.}~\bibnamefont {Uys}}, \ and\ \bibinfo
  {author} {\bibfnamefont {M.~J.}\ \bibnamefont {Biercuk}},\ }\href {\doibase
  10.1088/1367-2630/15/9/095004} {\bibfield  {journal} {\bibinfo  {journal}
  {New Journal of Physics}\ }\textbf {\bibinfo {volume} {15}},\ \bibinfo
  {pages} {095004} (\bibinfo {year} {2013})}\BibitemShut {NoStop}%
\bibitem [{\citenamefont {G\"ung\"ord\"u}\ and\ \citenamefont
  {Kestner}(2018)}]{Gungordu2018}%
  \BibitemOpen
  \bibfield  {author} {\bibinfo {author} {\bibfnamefont {U.}~\bibnamefont
  {G\"ung\"ord\"u}}\ and\ \bibinfo {author} {\bibfnamefont {J.~P.}\
  \bibnamefont {Kestner}},\ }\href {\doibase 10.1103/PhysRevB.98.165301}
  {\bibfield  {journal} {\bibinfo  {journal} {Phys. Rev. B}\ }\textbf {\bibinfo
  {volume} {98}},\ \bibinfo {pages} {165301} (\bibinfo {year}
  {2018})}\BibitemShut {NoStop}%
\bibitem [{\citenamefont {Ball}\ \emph {et~al.}(2020)\citenamefont {Ball},
  \citenamefont {Biercuk}, \citenamefont {Carvalho}, \citenamefont {Chen},
  \citenamefont {Hush}, \citenamefont {Castro}, \citenamefont {Li},
  \citenamefont {Liebermann}, \citenamefont {Slatyer}, \citenamefont {Edmunds},
  \citenamefont {Frey}, \citenamefont {Hempel},\ and\ \citenamefont
  {Milne}}]{Ball2020}%
  \BibitemOpen
  \bibfield  {author} {\bibinfo {author} {\bibfnamefont {H.}~\bibnamefont
  {Ball}}, \bibinfo {author} {\bibfnamefont {M.~J.}\ \bibnamefont {Biercuk}},
  \bibinfo {author} {\bibfnamefont {A.}~\bibnamefont {Carvalho}}, \bibinfo
  {author} {\bibfnamefont {J.}~\bibnamefont {Chen}}, \bibinfo {author}
  {\bibfnamefont {M.}~\bibnamefont {Hush}}, \bibinfo {author} {\bibfnamefont
  {L.~A.~D.}\ \bibnamefont {Castro}}, \bibinfo {author} {\bibfnamefont
  {L.}~\bibnamefont {Li}}, \bibinfo {author} {\bibfnamefont {P.~J.}\
  \bibnamefont {Liebermann}}, \bibinfo {author} {\bibfnamefont {H.~J.}\
  \bibnamefont {Slatyer}}, \bibinfo {author} {\bibfnamefont {C.}~\bibnamefont
  {Edmunds}}, \bibinfo {author} {\bibfnamefont {V.}~\bibnamefont {Frey}},
  \bibinfo {author} {\bibfnamefont {C.}~\bibnamefont {Hempel}}, \ and\ \bibinfo
  {author} {\bibfnamefont {A.}~\bibnamefont {Milne}},\ }\href@noop {} {\enquote
  {\bibinfo {title} {{Software tools for quantum control: Improving quantum
  computer performance through noise and error suppression}},}\ } (\bibinfo
  {year} {2020}),\ \Eprint {http://arxiv.org/abs/2001.04060} {arXiv:2001.04060}
  \BibitemShut {NoStop}%
\bibitem [{\citenamefont {van Dijk}\ \emph {et~al.}(2019)\citenamefont {van
  Dijk}, \citenamefont {Kawakami}, \citenamefont {Schouten}, \citenamefont
  {Veldhorst}, \citenamefont {Vandersypen}, \citenamefont {Babaie},
  \citenamefont {Charbon},\ and\ \citenamefont {Sebastiano}}]{Vandijk2018}%
  \BibitemOpen
  \bibfield  {author} {\bibinfo {author} {\bibfnamefont {J.}~\bibnamefont {van
  Dijk}}, \bibinfo {author} {\bibfnamefont {E.}~\bibnamefont {Kawakami}},
  \bibinfo {author} {\bibfnamefont {R.}~\bibnamefont {Schouten}}, \bibinfo
  {author} {\bibfnamefont {M.}~\bibnamefont {Veldhorst}}, \bibinfo {author}
  {\bibfnamefont {L.}~\bibnamefont {Vandersypen}}, \bibinfo {author}
  {\bibfnamefont {M.}~\bibnamefont {Babaie}}, \bibinfo {author} {\bibfnamefont
  {E.}~\bibnamefont {Charbon}}, \ and\ \bibinfo {author} {\bibfnamefont
  {F.}~\bibnamefont {Sebastiano}},\ }\href {\doibase
  10.1103/PhysRevApplied.12.044054} {\bibfield  {journal} {\bibinfo  {journal}
  {Phys. Rev. Applied}\ }\textbf {\bibinfo {volume} {12}},\ \bibinfo {pages}
  {044054} (\bibinfo {year} {2019})}\BibitemShut {NoStop}%
\bibitem [{\citenamefont {Cerfontaine}\ \emph {et~al.}(2014)\citenamefont
  {Cerfontaine}, \citenamefont {Botzem}, \citenamefont {DiVincenzo},\ and\
  \citenamefont {Bluhm}}]{Cerfontaine2014}%
  \BibitemOpen
  \bibfield  {author} {\bibinfo {author} {\bibfnamefont {P.}~\bibnamefont
  {Cerfontaine}}, \bibinfo {author} {\bibfnamefont {T.}~\bibnamefont {Botzem}},
  \bibinfo {author} {\bibfnamefont {D.~P.}\ \bibnamefont {DiVincenzo}}, \ and\
  \bibinfo {author} {\bibfnamefont {H.}~\bibnamefont {Bluhm}},\ }\href
  {\doibase 10.1103/PhysRevLett.113.150501} {\bibfield  {journal} {\bibinfo
  {journal} {Phys. Rev. Lett.}\ }\textbf {\bibinfo {volume} {113}},\ \bibinfo
  {pages} {150501} (\bibinfo {year} {2014})}\BibitemShut {NoStop}%
\bibitem [{\citenamefont {Khodjasteh}\ \emph {et~al.}(2013)\citenamefont
  {Khodjasteh}, \citenamefont {Sastrawan}, \citenamefont {Hayes}, \citenamefont
  {Green}, \citenamefont {Biercuk},\ and\ \citenamefont
  {Viola}}]{Khodjasteh2013}%
  \BibitemOpen
  \bibfield  {author} {\bibinfo {author} {\bibfnamefont {K.}~\bibnamefont
  {Khodjasteh}}, \bibinfo {author} {\bibfnamefont {J.}~\bibnamefont
  {Sastrawan}}, \bibinfo {author} {\bibfnamefont {D.}~\bibnamefont {Hayes}},
  \bibinfo {author} {\bibfnamefont {T.~J.}\ \bibnamefont {Green}}, \bibinfo
  {author} {\bibfnamefont {M.~J.}\ \bibnamefont {Biercuk}}, \ and\ \bibinfo
  {author} {\bibfnamefont {L.}~\bibnamefont {Viola}},\ }\href
  {https://www.ncbi.nlm.nih.gov/pubmed/23784079} {\bibfield  {journal}
  {\bibinfo  {journal} {Nature communications}\ }\textbf {\bibinfo {volume}
  {4}},\ \bibinfo {pages} {2045} (\bibinfo {year} {2013})}\BibitemShut
  {NoStop}%
\bibitem [{\citenamefont {Huang}\ and\ \citenamefont {Goan}(2017)}]{Huang2017}%
  \BibitemOpen
  \bibfield  {author} {\bibinfo {author} {\bibfnamefont {C.-H.}\ \bibnamefont
  {Huang}}\ and\ \bibinfo {author} {\bibfnamefont {H.-S.}\ \bibnamefont
  {Goan}},\ }\href {\doibase 10.1103/PhysRevA.95.062325} {\bibfield  {journal}
  {\bibinfo  {journal} {Phys. Rev. A}\ }\textbf {\bibinfo {volume} {95}},\
  \bibinfo {pages} {062325} (\bibinfo {year} {2017})}\BibitemShut {NoStop}%
\bibitem [{\citenamefont {{Hangleiter, Tobias and Cerfontaine, Pascal and
  Bluhm, Hendrik}}()}]{Hangleiter2021}%
  \BibitemOpen
  \bibfield  {author} {\bibinfo {author} {\bibnamefont {{Hangleiter, Tobias and
  Cerfontaine, Pascal and Bluhm, Hendrik}}},\ }\href@noop {} {\enquote
  {\bibinfo {title} {{Filter Function Formalism and Software Package to Compute
  Quantum Processes of Gate Sequences for Classical Non-Markovian Noise}},}\
  }\bibinfo {note} {{[Reference inserted by publisher]}}\BibitemShut {NoStop}%
\bibitem [{\citenamefont {Hangleiter}\ \emph {et~al.}(2021)\citenamefont
  {Hangleiter}, \citenamefont {Le},\ and\ \citenamefont {Teske}}]{software}%
  \BibitemOpen
  \bibfield  {author} {\bibinfo {author} {\bibfnamefont {T.}~\bibnamefont
  {Hangleiter}}, \bibinfo {author} {\bibfnamefont {I.~N.~M.}\ \bibnamefont
  {Le}}, \ and\ \bibinfo {author} {\bibfnamefont {J.~D.}\ \bibnamefont
  {Teske}},\ }\href {\doibase 10.5281/zenodo.4575000} {\enquote {\bibinfo
  {title} {{filter\_functions: A package for efficient numerical calculation of
  generalized filter functions to describe the effect of noise on quantum gate
  operations}},}\ } (\bibinfo {year} {2021}),\ \bibinfo {note} {available at
  \url{https://github.com/qutech/filter_functions/}}\BibitemShut {NoStop}%
\bibitem [{\citenamefont {Magnus}(1954)}]{Magnus1954}%
  \BibitemOpen
  \bibfield  {author} {\bibinfo {author} {\bibfnamefont {W.}~\bibnamefont
  {Magnus}},\ }\href {\doibase 10.1002/cpa.3160070404} {\bibfield  {journal}
  {\bibinfo  {journal} {{Communications on Pure and Applied Mathematics}}\
  }\textbf {\bibinfo {volume} {7}},\ \bibinfo {pages} {649} (\bibinfo {year}
  {1954})}\BibitemShut {NoStop}%
\bibitem [{\citenamefont {Blanes}\ \emph {et~al.}(2009)\citenamefont {Blanes},
  \citenamefont {Casas}, \citenamefont {Oteo},\ and\ \citenamefont
  {Ros}}]{Blanes2009}%
  \BibitemOpen
  \bibfield  {author} {\bibinfo {author} {\bibfnamefont {S.}~\bibnamefont
  {Blanes}}, \bibinfo {author} {\bibfnamefont {F.}~\bibnamefont {Casas}},
  \bibinfo {author} {\bibfnamefont {J.}~\bibnamefont {Oteo}}, \ and\ \bibinfo
  {author} {\bibfnamefont {J.}~\bibnamefont {Ros}},\ }\href {\doibase
  10.1016/j.physrep.2008.11.001} {\bibfield  {journal} {\bibinfo  {journal}
  {Physics Reports}\ }\textbf {\bibinfo {volume} {470}},\ \bibinfo {pages}
  {151} (\bibinfo {year} {2009})}\BibitemShut {NoStop}%
\bibitem [{Note1()}]{Note1}%
  \BibitemOpen
  \bibinfo {note} {While $\xi $ given here is only valid for time-independent
  $\protect \ensuremath {B_\alpha }\protect \xspace $, an extended discussion
  of the convergence criteria is given in Ref.~\protect \citenum
  {Hangleiter2021} and Ref.~\protect \citenum {Green2013}.}\BibitemShut {Stop}%
\bibitem [{\citenamefont {Haeberlen}(1976)}]{Haeberlen1976}%
  \BibitemOpen
  \bibfield  {author} {\bibinfo {author} {\bibfnamefont {U.}~\bibnamefont
  {Haeberlen}},\ }\href {\doibase 10.1016/B978-0-12-025561-0.X5001-1} {\emph
  {\bibinfo {title} {Advances in Magnetic Resonance}}},\ edited by\ \bibinfo
  {editor} {\bibfnamefont {J.~S.}\ \bibnamefont {Waugh}}\ (\bibinfo
  {publisher} {Elsevier},\ \bibinfo {address} {New York},\ \bibinfo {year}
  {1976})\BibitemShut {NoStop}%
\bibitem [{\citenamefont {Majenz}\ \emph {et~al.}(2013)\citenamefont {Majenz},
  \citenamefont {Albash}, \citenamefont {Breuer},\ and\ \citenamefont
  {Lidar}}]{Majenz2013}%
  \BibitemOpen
  \bibfield  {author} {\bibinfo {author} {\bibfnamefont {C.}~\bibnamefont
  {Majenz}}, \bibinfo {author} {\bibfnamefont {T.}~\bibnamefont {Albash}},
  \bibinfo {author} {\bibfnamefont {H.-P.}\ \bibnamefont {Breuer}}, \ and\
  \bibinfo {author} {\bibfnamefont {D.~A.}\ \bibnamefont {Lidar}},\ }\href
  {\doibase 10.1103/PhysRevA.88.012103} {\bibfield  {journal} {\bibinfo
  {journal} {Physical Review A}\ }\textbf {\bibinfo {volume} {88}},\ \bibinfo
  {pages} {012103} (\bibinfo {year} {2013})}\BibitemShut {NoStop}%
\bibitem [{\citenamefont {Cerfontaine}\ \emph {et~al.}(2020)\citenamefont
  {Cerfontaine}, \citenamefont {Otten},\ and\ \citenamefont
  {Bluhm}}]{Cerfontaine2019gsc}%
  \BibitemOpen
  \bibfield  {author} {\bibinfo {author} {\bibfnamefont {P.}~\bibnamefont
  {Cerfontaine}}, \bibinfo {author} {\bibfnamefont {R.}~\bibnamefont {Otten}},
  \ and\ \bibinfo {author} {\bibfnamefont {H.}~\bibnamefont {Bluhm}},\ }\href
  {\doibase 10.1103/PhysRevApplied.13.044071} {\bibfield  {journal} {\bibinfo
  {journal} {Phys. Rev. Applied}\ }\textbf {\bibinfo {volume} {13}},\ \bibinfo
  {pages} {044071} (\bibinfo {year} {2020})}\BibitemShut {NoStop}%
\bibitem [{\citenamefont {Kimmel}\ \emph {et~al.}(2015)\citenamefont {Kimmel},
  \citenamefont {Low},\ and\ \citenamefont {Yoder}}]{Kimmel2015}%
  \BibitemOpen
  \bibfield  {author} {\bibinfo {author} {\bibfnamefont {S.}~\bibnamefont
  {Kimmel}}, \bibinfo {author} {\bibfnamefont {G.~H.}\ \bibnamefont {Low}}, \
  and\ \bibinfo {author} {\bibfnamefont {T.~J.}\ \bibnamefont {Yoder}},\ }\href
  {\doibase 10.1103/PhysRevA.92.062315} {\bibfield  {journal} {\bibinfo
  {journal} {Phys. Rev. A}\ }\textbf {\bibinfo {volume} {92}},\ \bibinfo
  {pages} {062315} (\bibinfo {year} {2015})}\BibitemShut {NoStop}%
\bibitem [{\citenamefont {Reed}(2013)}]{Reed2013}%
  \BibitemOpen
  \bibfield  {author} {\bibinfo {author} {\bibfnamefont {M.}~\bibnamefont
  {Reed}},\ }\emph {\bibinfo {title} {{Entanglement and Quantum Error
  Correction with Superconducting Qubits}}},\ \href
  {http://arxiv.org/abs/1311.6759} {\bibinfo {type} {Phd thesis}},\ \bibinfo
  {school} {{Yale University}} (\bibinfo {year} {2013}),\ \Eprint
  {http://arxiv.org/abs/1311.6759} {arXiv:1311.6759} \BibitemShut {NoStop}%
\bibitem [{\citenamefont {Kelly}\ \emph {et~al.}(2014)\citenamefont {Kelly},
  \citenamefont {Barends}, \citenamefont {Campbell}, \citenamefont {Chen},
  \citenamefont {Chen}, \citenamefont {Chiaro}, \citenamefont {Dunsworth},
  \citenamefont {Fowler}, \citenamefont {Hoi}, \citenamefont {Jeffrey},
  \citenamefont {Megrant}, \citenamefont {Mutus}, \citenamefont {Neill},
  \citenamefont {O'Malley}, \citenamefont {Quintana}, \citenamefont {Roushan},
  \citenamefont {Sank}, \citenamefont {Vainsencher}, \citenamefont {Wenner},
  \citenamefont {White}, \citenamefont {Cleland},\ and\ \citenamefont
  {Martinis}}]{Kelly2014}%
  \BibitemOpen
  \bibfield  {author} {\bibinfo {author} {\bibfnamefont {J.}~\bibnamefont
  {Kelly}}, \bibinfo {author} {\bibfnamefont {R.}~\bibnamefont {Barends}},
  \bibinfo {author} {\bibfnamefont {B.}~\bibnamefont {Campbell}}, \bibinfo
  {author} {\bibfnamefont {Y.}~\bibnamefont {Chen}}, \bibinfo {author}
  {\bibfnamefont {Z.}~\bibnamefont {Chen}}, \bibinfo {author} {\bibfnamefont
  {B.}~\bibnamefont {Chiaro}}, \bibinfo {author} {\bibfnamefont
  {A.}~\bibnamefont {Dunsworth}}, \bibinfo {author} {\bibfnamefont {A.~G.}\
  \bibnamefont {Fowler}}, \bibinfo {author} {\bibfnamefont {I.-C.}\
  \bibnamefont {Hoi}}, \bibinfo {author} {\bibfnamefont {E.}~\bibnamefont
  {Jeffrey}}, \bibinfo {author} {\bibfnamefont {A.}~\bibnamefont {Megrant}},
  \bibinfo {author} {\bibfnamefont {J.}~\bibnamefont {Mutus}}, \bibinfo
  {author} {\bibfnamefont {C.}~\bibnamefont {Neill}}, \bibinfo {author}
  {\bibfnamefont {P.~J.~J.}\ \bibnamefont {O'Malley}}, \bibinfo {author}
  {\bibfnamefont {C.}~\bibnamefont {Quintana}}, \bibinfo {author}
  {\bibfnamefont {P.}~\bibnamefont {Roushan}}, \bibinfo {author} {\bibfnamefont
  {D.}~\bibnamefont {Sank}}, \bibinfo {author} {\bibfnamefont {A.}~\bibnamefont
  {Vainsencher}}, \bibinfo {author} {\bibfnamefont {J.}~\bibnamefont {Wenner}},
  \bibinfo {author} {\bibfnamefont {T.~C.}\ \bibnamefont {White}}, \bibinfo
  {author} {\bibfnamefont {A.~N.}\ \bibnamefont {Cleland}}, \ and\ \bibinfo
  {author} {\bibfnamefont {J.~M.}\ \bibnamefont {Martinis}},\ }\href {\doibase
  10.1103/PhysRevLett.112.240504} {\bibfield  {journal} {\bibinfo  {journal}
  {Physical Review Letters}\ }\textbf {\bibinfo {volume} {112}},\ \bibinfo
  {pages} {240504} (\bibinfo {year} {2014})}\BibitemShut {NoStop}%
\bibitem [{\citenamefont {Egger}\ and\ \citenamefont
  {Wilhelm}(2014)}]{Egger2014}%
  \BibitemOpen
  \bibfield  {author} {\bibinfo {author} {\bibfnamefont {D.~J.}\ \bibnamefont
  {Egger}}\ and\ \bibinfo {author} {\bibfnamefont {F.~K.}\ \bibnamefont
  {Wilhelm}},\ }\href {\doibase 10.1103/PhysRevLett.112.240503} {\bibfield
  {journal} {\bibinfo  {journal} {Physical Review Letters}\ }\textbf {\bibinfo
  {volume} {112}},\ \bibinfo {pages} {240503} (\bibinfo {year}
  {2014})}\BibitemShut {NoStop}%
\bibitem [{Note2()}]{Note2}%
  \BibitemOpen
  \bibinfo {note} {Note that this affects all perturbative approaches, e.g.\
  Ref.~\protect \citenum {Green2013}}\BibitemShut {NoStop}%
\bibitem [{\citenamefont {Choi}(1975)}]{Choi1975}%
  \BibitemOpen
  \bibfield  {author} {\bibinfo {author} {\bibfnamefont {M.-D.}\ \bibnamefont
  {Choi}},\ }\href {\doibase 10.1016/0024-3795(75)90075-0} {\bibfield
  {journal} {\bibinfo  {journal} {Linear Algebra Appl.}\ }\textbf {\bibinfo
  {volume} {10}},\ \bibinfo {pages} {285} (\bibinfo {year} {1975})}\BibitemShut
  {NoStop}%
\bibitem [{\citenamefont {Hall}\ \emph {et~al.}(2014)\citenamefont {Hall},
  \citenamefont {Cresser}, \citenamefont {Li},\ and\ \citenamefont
  {Andersson}}]{Hall2014}%
  \BibitemOpen
  \bibfield  {author} {\bibinfo {author} {\bibfnamefont {M.~J.~W.}\
  \bibnamefont {Hall}}, \bibinfo {author} {\bibfnamefont {J.~D.}\ \bibnamefont
  {Cresser}}, \bibinfo {author} {\bibfnamefont {L.}~\bibnamefont {Li}}, \ and\
  \bibinfo {author} {\bibfnamefont {E.}~\bibnamefont {Andersson}},\ }\href
  {\doibase 10.1103/PhysRevA.89.042120} {\bibfield  {journal} {\bibinfo
  {journal} {Phys. Rev. A}\ }\textbf {\bibinfo {volume} {89}},\ \bibinfo
  {pages} {042120} (\bibinfo {year} {2014})}\BibitemShut {NoStop}%
\bibitem [{Note3()}]{Note3}%
  \BibitemOpen
  \bibinfo {note} {See Supplemental Material \cite {prlSupp} for a brief
  overview, which includes Refs. \protect \citenum
  {Wood2018,Wallman2016,Johansson2013,Schulte-herbruggen2005,Gorini1976}}\BibitemShut
  {NoStop}%
\bibitem [{\citenamefont {Nielsen}(2002)}]{Nielsen2002}%
  \BibitemOpen
  \bibfield  {author} {\bibinfo {author} {\bibfnamefont {M.~A.}\ \bibnamefont
  {Nielsen}},\ }\href {\doibase 10.1016/S0375-9601(02)01272-0} {\bibfield
  {journal} {\bibinfo  {journal} {Physics Letters A}\ }\textbf {\bibinfo
  {volume} {303}},\ \bibinfo {pages} {249} (\bibinfo {year}
  {2002})}\BibitemShut {NoStop}%
\bibitem [{\citenamefont {Kimmel}\ \emph {et~al.}(2014)\citenamefont {Kimmel},
  \citenamefont {da~Silva}, \citenamefont {Ryan}, \citenamefont {Johnson},\
  and\ \citenamefont {Ohki}}]{Kimmel2014}%
  \BibitemOpen
  \bibfield  {author} {\bibinfo {author} {\bibfnamefont {S.}~\bibnamefont
  {Kimmel}}, \bibinfo {author} {\bibfnamefont {M.~P.}\ \bibnamefont
  {da~Silva}}, \bibinfo {author} {\bibfnamefont {C.~A.}\ \bibnamefont {Ryan}},
  \bibinfo {author} {\bibfnamefont {B.~R.}\ \bibnamefont {Johnson}}, \ and\
  \bibinfo {author} {\bibfnamefont {T.}~\bibnamefont {Ohki}},\ }\href {\doibase
  10.1103/PhysRevX.4.011050} {\bibfield  {journal} {\bibinfo  {journal} {Phys.
  Rev. X}\ }\textbf {\bibinfo {volume} {4}},\ \bibinfo {pages} {011050}
  (\bibinfo {year} {2014})}\BibitemShut {NoStop}%
\bibitem [{Note4()}]{Note4}%
  \BibitemOpen
  \bibinfo {note} {$F_\alpha ^{(gg')}(\omega ) = \protect \ensuremath {\protect
  \ensuremath {\protect \mathrm {e}}\protect \xspace }^{\protect \ensuremath
  {\protect \ensuremath {\protect \mathrm {i}}\protect \xspace }\omega (t_{g-1}
  - t_{g'-1})}\times \\\protect \ensuremath {\protect \mathrm {tr}}\protect
  \bigl (\protect \ensuremath {\protect \mathcal {Q}}\protect \xspace \protect
  \ensuremath {^{(g'-1)}}\protect \xspace \protect \bigl (\protect \ensuremath
  {\protect \tilde {B}_\alpha }\protect \xspace \protect \ensuremath
  {^{(g')}}\protect \xspace (\omega )\protect \bigr )\protect \ensuremath
  {^\dagger }\protect \xspace \protect \ensuremath {\protect \mathcal
  {Q}}\protect \xspace \protect \ensuremath {^{(g-1)}}\protect \xspace \protect
  \bigl (\protect \ensuremath {\protect \tilde {B}_\alpha }\protect \xspace
  \protect \ensuremath {^{(g)}}\protect \xspace (\omega )\protect \bigr
  )\protect \bigr )$}\BibitemShut {NoStop}%
\bibitem [{\citenamefont {Coppersmith}(2002)}]{Coppersmith1994}%
  \BibitemOpen
  \bibfield  {author} {\bibinfo {author} {\bibfnamefont {D.}~\bibnamefont
  {Coppersmith}},\ }\href@noop {} {\enquote {\bibinfo {title} {{An approximate
  Fourier transform useful in quantum factoring}},}\ } (\bibinfo {year}
  {2002}),\ \Eprint {http://arxiv.org/abs/quant-ph/0201067}
  {arXiv:quant-ph/0201067} \BibitemShut {NoStop}%
\bibitem [{Note5()}]{Note5}%
  \BibitemOpen
  \bibinfo {note} {Example code available at \protect \url
  {https://github.com/qutech/filter_functions/blob/master/doc/source/examples/quantum_fourier_transform.ipynb}}\BibitemShut
  {NoStop}%
\bibitem [{Note6()}]{Note6}%
  \BibitemOpen
  \bibinfo {note} {See Supplemental Material \cite {prlSupp} for the circuit
  diagram and the physical model}\BibitemShut {NoStop}%
\bibitem [{\citenamefont {Le}\ \emph {et~al.}()\citenamefont {Le},
  \citenamefont {Teske}, \citenamefont {Hangleiter}, \citenamefont
  {Cerfontaine},\ and\ \citenamefont {Bluhm}}]{Le2021}%
  \BibitemOpen
  \bibfield  {author} {\bibinfo {author} {\bibfnamefont {I.~N.~M.}\
  \bibnamefont {Le}}, \bibinfo {author} {\bibfnamefont {J.~D.}\ \bibnamefont
  {Teske}}, \bibinfo {author} {\bibfnamefont {T.}~\bibnamefont {Hangleiter}},
  \bibinfo {author} {\bibfnamefont {P.}~\bibnamefont {Cerfontaine}}, \ and\
  \bibinfo {author} {\bibfnamefont {H.}~\bibnamefont {Bluhm}},\ }\href
  {http://arxiv.org/abs/2103.09126} {\enquote {\bibinfo {title} {{Analytic
  Filter Function Derivatives for Quantum Optimal Control}},}\ }\Eprint
  {http://arxiv.org/abs/2103.09126} {arXiv:2103.09126} \BibitemShut {NoStop}%
\bibitem [{prl()}]{prlSupp}%
  \BibitemOpen
  \href@noop {} {}\bibinfo {note} {[URL inserted by publisher]}\BibitemShut
  {NoStop}%
\bibitem [{\citenamefont {Wood}\ and\ \citenamefont
  {Gambetta}(2018)}]{Wood2018}%
  \BibitemOpen
  \bibfield  {author} {\bibinfo {author} {\bibfnamefont {C.~J.}\ \bibnamefont
  {Wood}}\ and\ \bibinfo {author} {\bibfnamefont {J.~M.}\ \bibnamefont
  {Gambetta}},\ }\href {\doibase 10.1103/PhysRevA.97.032306} {\bibfield
  {journal} {\bibinfo  {journal} {Phys. Rev. A}\ }\textbf {\bibinfo {volume}
  {97}},\ \bibinfo {pages} {032306} (\bibinfo {year} {2018})}\BibitemShut
  {NoStop}%
\bibitem [{\citenamefont {Sanders}\ \emph {et~al.}(2015)\citenamefont
  {Sanders}, \citenamefont {Wallman},\ and\ \citenamefont
  {Sanders}}]{Wallman2016}%
  \BibitemOpen
  \bibfield  {author} {\bibinfo {author} {\bibfnamefont {Y.~R.}\ \bibnamefont
  {Sanders}}, \bibinfo {author} {\bibfnamefont {J.~J.}\ \bibnamefont
  {Wallman}}, \ and\ \bibinfo {author} {\bibfnamefont {B.~C.}\ \bibnamefont
  {Sanders}},\ }\href
  {https://stacks.iop.org/1367-2630/18/i=1/a=012002?key=crossref.50c14015cc7be50527fcc46b24b602d3}
  {\bibfield  {journal} {\bibinfo  {journal} {New Journal of Physics}\ }\textbf
  {\bibinfo {volume} {18}},\ \bibinfo {pages} {012002} (\bibinfo {year}
  {2015})}\BibitemShut {NoStop}%
\bibitem [{\citenamefont {Johansson}\ \emph {et~al.}(2013)\citenamefont
  {Johansson}, \citenamefont {Nation},\ and\ \citenamefont
  {Nori}}]{Johansson2013}%
  \BibitemOpen
  \bibfield  {author} {\bibinfo {author} {\bibfnamefont {J.}~\bibnamefont
  {Johansson}}, \bibinfo {author} {\bibfnamefont {P.}~\bibnamefont {Nation}}, \
  and\ \bibinfo {author} {\bibfnamefont {F.}~\bibnamefont {Nori}},\ }\href
  {\doibase 10.1016/j.cpc.2012.11.019} {\bibfield  {journal} {\bibinfo
  {journal} {Computer Physics Communications}\ }\textbf {\bibinfo {volume}
  {184}},\ \bibinfo {pages} {1234} (\bibinfo {year} {2013})}\BibitemShut
  {NoStop}%
\bibitem [{\citenamefont {Schulte-Herbr{\"{u}}ggen}\ \emph
  {et~al.}(2005)\citenamefont {Schulte-Herbr{\"{u}}ggen}, \citenamefont
  {Sp{\"{o}}rl}, \citenamefont {Khaneja},\ and\ \citenamefont
  {Glaser}}]{Schulte-herbruggen2005}%
  \BibitemOpen
  \bibfield  {author} {\bibinfo {author} {\bibfnamefont {T.}~\bibnamefont
  {Schulte-Herbr{\"{u}}ggen}}, \bibinfo {author} {\bibfnamefont
  {A.}~\bibnamefont {Sp{\"{o}}rl}}, \bibinfo {author} {\bibfnamefont
  {N.}~\bibnamefont {Khaneja}}, \ and\ \bibinfo {author} {\bibfnamefont
  {S.~J.}\ \bibnamefont {Glaser}},\ }\href {\doibase
  10.1103/PhysRevA.72.042331} {\bibfield  {journal} {\bibinfo  {journal} {Phys.
  Rev. A}\ }\textbf {\bibinfo {volume} {72}},\ \bibinfo {pages} {042331}
  (\bibinfo {year} {2005})}\BibitemShut {NoStop}%
\bibitem [{\citenamefont {Gorini}\ \emph {et~al.}(1976)\citenamefont {Gorini},
  \citenamefont {Kossakowski},\ and\ \citenamefont {Sudarshan}}]{Gorini1976}%
  \BibitemOpen
  \bibfield  {author} {\bibinfo {author} {\bibfnamefont {V.}~\bibnamefont
  {Gorini}}, \bibinfo {author} {\bibfnamefont {A.}~\bibnamefont {Kossakowski}},
  \ and\ \bibinfo {author} {\bibfnamefont {E.~C.~G.}\ \bibnamefont
  {Sudarshan}},\ }\href {\doibase 10.1063/1.522979} {\bibfield  {journal}
  {\bibinfo  {journal} {Journal of Mathematical Physics}\ }\textbf {\bibinfo
  {volume} {17}},\ \bibinfo {pages} {821} (\bibinfo {year} {1976})}\BibitemShut
  {NoStop}%
\end{thebibliography}
\end{document}